\documentclass[letterpaper]{article}
\usepackage{amsmath,amssymb,graphicx,enumerate,footnote}
\usepackage{verbatim,graphicx}
\usepackage{graphicx,color,verbatim,enumerate}
\usepackage{caption,subcaption}%
\begin{document}
\title{A Novel Method to Calculate Click Through Rate for Sponsored Search}
\author{Rahul Gupta\thanks{This work was carried out at MIT, Manipal. Currently Rahul Gupta is working as Data Scientist at Fractal Analytics Inc, India, 
Email: rahul.gupta.engg@gmail.com}, Gitansh Khirbat \thanks{Gitansh Khirbat is at University of Melbourne, Australia, Email:gitanshkhirbat.91@gmail.com} and Sanjay Singh\thanks{Sanjay Singh is with the Department of Information and Communication Technology, Manipal Institute of Technology, Manipal University, Manipal-576104, INDIA, E-mail: sanjay.singh@manipal.edu}}

\maketitle              
\begin{abstract}        
Sponsored search adopts generalized second price (GSP) auction mechanism which works on the concept of pay per click which is most commonly used for the allocation of slots in the searched page. Two main aspects associated with GSP are the bidding amount and the click through rate (CTR). The CTR learning algorithms currently being used works on the basic principle of ($\#\mbox{clicks}_i/ \#\mbox{impressions}_i$) under a fixed window of clicks or impressions or time. CTR are prone to fraudulent clicks, resulting in sudden increase of CTR. The current algorithms are unable to find the solutions to stop this, although with the use of machine learning algorithms it can be detected that fraudulent clicks are being generated. In our paper, we have used the concept of relative ranking which works on the basic principle of ($\#\mbox{clicks}_i /\#\mbox{clicks}_t$). In this algorithm, both the numerator and the denominator are linked. As $\#\mbox{clicks}_t$ is higher than previous algorithms and is linked to the $\#\mbox{clicks}_i$, the small change in the clicks which occurs in the normal scenario have a very small change in the result but in case of fraudulent clicks the number of clicks increases or decreases rapidly which will add up with the normal clicks to increase the denominator, thereby decreasing the CTR. 
\end{abstract}

\section{Introduction}
The sponsored search algorithms are the most vital aspect of any search engine as the sponsored ads are its main source of income. The free service that the search engines are providing is being possible because of the sponsored advertisements that appear on the web page whenever any commercial words are search. So each keyword is associated with a group of advertisers who are ready to bid for it. Ideally the slots available to display those ads are less than the number of bidders that are bidding for it. In 2011, Google ad-words resulted an income of \$37.9 billion which accounts for 96\% of its total income \cite{abc}. The keyword ranges from Finance \& Insurance, Retailers \& Merchandise, Travel \& Tourism, Jobs and Education Home \& Garden, Computer \& Consumer Electronics, Vehicles, Internet \& Telecoms, Business \& Industrial, Occasions \& Gifts with companies some companies paying as high as \$50 million \cite{ghi} \cite{hij}.  The same is the case of other search engine like yahoo too, out of \$2.02 billion revenue generated in 2011, display ads constitutes \$1.2 billion which is equivalent to 61\% of its total income \cite{ijk}. So, display-ad is the most important part of any search engine.
%

The sole motivation of taking this work is an attempt to learn how are the search engines able earn a very high profit and maintain top most rank in the fortune 100 companies year after year even though they are not charging from the users. Each year their earnings are growing exponentially, which accounts for the popularity these ads holds. Currently, there are more than 65,000 keywords for which real time auction goes on among 800,000 bidders \cite{jkl}. The most interesting feature is that this profit is unaffected by any political, social, economical aspects. People will be using the search engine for searching even though there is a hike in petrol price or a tax rebate. So the whole idea reflects that this is a business which will always be growing.

\section{Related Work}
This section briefs about the existing work in the area of sponsored search.

\subsection{Pay Per Impression}
In 1994, the first time ads appeared online. The first ad that appeared online was of AT\&T appeared at Hotmail.com which works on the principle of pay per click. Cost per impression is used in online advertising and marketing related to web traffic. It refers to the cost of Internet marketing campaigns where advertisers pay for every time their ad is displayed, usually in the form of a banner ad on in Email advertising. An impression is the display of an ad to a user while viewing a web page. A single web page may contain multiple ads. In such cases, a single page-view would result in one impression for each ad displayed. 
\par
In order to count the impressions accurately and prevent fraud, an ad server may exclude certain non-qualifying activities such as page-refreshes or the user opening same pages from counting as impressions \cite{xyz}. 

\subsection{Pay Per Acquisition}
In 1996, companies like Amazon, CDNow used the concept of pay per acquisition. According to which, the advertiser pays for each specified action (a purchase, a form submission, and so on) linked to the advertisement. Direct response advertisers consider CPA the optimal way to buy online advertising, as an advertiser only pays for the ad when the desired action has occurred. An action can be a product being purchased, a form being filled, etc. The desired action to be performed is determined by the advertiser. A merchant would team up with other merchants or partner websites to promote its products or services online. The merchant would only pay out when confirmed lead, sale, email or registration takes place. Ideally, an advertiser would always prefer a pricing model in which the advertiser pays only when a customer actually completes a transaction. The PPT (pay per transaction) models were born out of this contention. A prominent example of PPT models is Amazon.com's Associates Program. Under this program, a website that sends customers to Amazon.com receives a percentage of customer's purchases.
\par
This mechanism is highly beneficial for the advertiser as they had to pay only when a particular task which they want is done by the user. But for the host website, this is not trustworthy. As when the user has left the site and is redirected to the other site, it is impossible to keep a track of activities that one performs on the site in which he was directed. So this method can only be adopted when the advertiser is highly trustworthy or he agrees to share all the data to its host site also in which its advertisement had appear.

\subsection{Pay Per Click}
This is the most popular mechanism which forms the basis for the sponsored ads. It was first time started by GoTo.com in 1998. In this method, the advertiser has to pay every time a user clicks on the ads held by the publisher website. It is a middle ware between PPI \& PPA as it provides a well trusted way for both the advertiser and the host website. 

\section{Pay Per Click with Generalized First Price Auction}
Till now all the web page considered for advertisement have a static framework which means that there is already a well defined number of pages and the web page displayed will be amongst them. But when advertisement in search engine comes into picture, the count of web page is not static as it displays different result based on the queries. So here for each keyword an auction is being is conducted where auctioneers bid for slots in the search engine; the number of slots associated with a keyword is lesser than the no of companies bidding for it.  In generalized first price auction, bidders place their bid in a sealed envelope and simultaneously hand them to the auctioneer. The envelopes are opened and the individual with the highest bid wins, paying a price equal to the exact amount that he or she bid. 
\par
This method was first adopted by Yahoo in 2001, where an online auction was conducted on a set of keywords \cite{bcd}. This method which became very popular in the beginning soon suffered from very high in congruence which is explained the next section.The number of ads that the search engine can show to a user is limited with different desirability for advertisers: an ad shown at the top of a page is more likely to be clicked than an ad shown at the bottom. In the search engine industry, there are three key players: the advertisers, the search engines and the potential customers. Search engines navigate potential customers to advertisers' product web sites by displaying their web links when potential customers conduct keyword search requests. These advertisers' links are called sponsored links. Sponsored links distinguish themselves from the organic (non-sponsored) web search results by whether or not a fee is paid to the search engine company. Every time a consumer clicked on a sponsored link, an advertiser's account was automatically billed the amount of the advertiser's most recent bid. The links to advertisers were arranged in descending order of bids, making highest bids the most prominent. The ease of use, the very low entry costs, and the transparency of the mechanism quickly led to the success of Overture's paid search platform.However, the underlying auction mechanism itself was far from perfect. In particular, Overture and advertisers quickly learned that the mechanism was unstable due to the dynamic nature of the environment.
\par
After just 6 months of the existence, GFP failed as it suffered from serious glitches. The auction in the search engine is dynamic which means that a user can any time change the amount of bid which he/she has bid. Due to the same, this system failed to work as the competition among them causes a instability. This could be well understood with the help of an example: Consider an example where we have 2 slots available for a particular keyword. There was a bidding being held for a particular keyword in which advertiser A with bid amount \$ 10 won the first slot and the advertiser B with the bidding amount of \$ 3 won the second slot. Now as we know that the companies can change their bidding amount at any instant. This prompts the advertiser A to change its bid to \$ 4 as with that bid too he can maintain its first position. Now the advertiser A holds the first spot with \$ 4   and the advertiser B holds the second spot with \$ 3. This attracts the advertiser B to increase its bid to \$ 5, so that it can acquire the first slot. This will again be followed by advertiser B increasing its bid to \$ 6 to regain its spot. This follows until a particular level is reached where advertiser B will stop the increase. And advertiser A will have the first position. Now at that instant advertiser B will suddenly lower its value which accounts for the least possible value to b in that slot. This will again result in the previous scenario making the advertiser A to lower its bid \cite{cde}.  

\section{Generalized Second Price Auction Mechanism}
This is the auction mechanism currently being followed by most of the search engines which includes Google, Yahoo, Bing. In this concept, each bidder pays the amount which is given by the next bidder following him \cite{efg}. The highest bidder pays the amount of the second highest and the second highest pays the amount bidded by the third highest, this goes on till the last available slot is allotted \cite{lmn}. The slot $S_i$ if given to an advertiser A, it has to pay a price $P_i=b_{i-1}$.

\subsection{Click Through Rate (CTR)}
Click Through Rate is the quantitative measure of the ads which defines the probability an ad will get the click if it is shown. Each ads has an individual CTR which is based on the prior experience that it holds in the particular search engine. It also denotes the popularity an ad has among the user. CTR is multiplied by the bid amount that the advertiser is ready to pay and the answer is the final bid amount for that advertisement. So higher CTR is considered better as it will increase the bid and as a result the advertiser will be able to get a better slot with less amount of payment.\par
Ideally if any of the advertisement wants its CTR to be increased, it should take care of the service it is providing. The better the service/user-interface, the user will be more prompt to visit that ad which will subsequently increase its CTR. But practically it is sometimes not being followed, fraudulent clicks are done to increase the CTR of their own ads or to increase the spending of other's ad. 

\section{Clicks Fraud}
Click Fraud is the act of clicking on the the ads with the sole intent of increasing or decreasing the spending of the advertiser. 
In a pay per click mechanism, the advertiser has to pay every times its ad is clicked. This makes the system prone to fraudulent clicks \cite{fgh}. 
\par
Clicking on other's ad- As every click resulted in the payment to be paid by the advertisement. If the rivalry company performs the fraudulent clicks to the other company's advertisement, it will result in the unnecessary payment to be paid by the other company. 
Clicking on own's ad- This is done in order to increase the CTR of own ads so that it can acquire the better slots. As it is general proven fact that the higher is the slot, higher will be the probability to click on that ad. Higher position in the slot not only gives more genuine clicks but also decrease the cost to be paid per click.

\section{Generating Click Fraud}
Click fraud is done by many ways which are broadly classified in two ways:

\subsection{ Automatically- by scripting}
Here, a script(code) is used to generate clicks on the particular advertisement. This leads to a sudden increase in the number of clicks thereby increasing the CTR. The implementation of this method is quite simple, as no external resources is being used. However this method can be detected by the search engines. This is done by a variety of steps that are developed by examining the patterns of clicks that are generated by scripted clicks. Like when a click is scripted each click will occur generally occur after a fixed interval of time generally in milliseconds. This is not possible in practical scenario, so whenever a similar patterns of clicks are encountered for a longer duration of time it is discarded. 

\subsection{By humans}
This is a way of generating fraudulent clicks where humans are employed to click on the particular ads. Here in this case an ads is attacked by the set people where they visit the ad shown and spend the time like a normal user do. As there  is no fixed pattern and their likelihood to behave like the normal user, it is almost impossible to detect the fraud done by this method.

\section{Existing CTR Learning Algorithm}
There is in general 3 CTR learning algorithms currently being used which works on the basic principle of (\#clicks $_{i}$/\#impressions $_{i}$) under a fixed window of clicks or impressions or time. CTR are prone to fraudulent clicks, resulting in the sudden increase of CTR. The current algorithms are unable to find the solutions to stop this.

\subsection {Averaging Over Fixed Time Window T}
Let X= No of clicks in last T hours and Y= No of impressions.
\begin{equation}
CTR= X/Y
\end{equation}

\subsection {Averaging over fixed impression window Y}
Let X= No of clicks received in last Y impression. 
\begin{equation}
CTR= X/Y
\end{equation}

\subsection {Averaging over fixed window X}
Let Y = impression since the Xth last click. 
\begin{equation}
CTR= X/Y
\end{equation}

\section{New CTR Learning Algorithm}
Let A,B,C be three advertisers and $C_a$, $C_b$, $C_c$ be the clicks that the advertisers receive in time interval $T$. 
\begin{eqnarray}
CTR_{A}= C_{a}/(C_{a} + C_{b}+ C_{c})\\
CTR_{B}= C_{b}/(C_{a}+C_{b}+C_{c})\\
CTR_{C}= C_{c}/(C_{a}+C_{b}+C_{c})
\end{eqnarray}
\section{Results and Discussion}
Below is the table and the graph obtained by the old algorithms and the new algorithms when the fraudulent clicks are being generated. The graph clearly depicts that the new algorithm is better then the previous one.

\begin{table}[bpht!]
\centering
\caption{Data for the calculation of CTR using current algorithm}
\label{tab:t1}
\begin{tabular}{|l|l|l|l|}
    \hline
    Time & Impressions & Clicks & CTR\\ \hline
    1 & 16 & 2 & 0.111 \\ \hline
    2 & 28 & 6 & 0.176 \\ \hline
    3 & 40 & 12 & 0.230 \\ \hline
		4 & 52 & 18 & 0.257 \\ \hline
    5 & 64 & 24 & 0.272 \\ \hline
    6 & 76 & 30 & 0.283 \\ \hline
		7 & 88 & 42 & 0.290 \\ \hline
    8 & 100 & 48 & 0.295 \\ \hline
    9 & 112 & 54 & 3.0 \\ \hline
		10 & 124 & 60 & 0.303 \\ \hline
    11 & 136 & 66 & 0.306 \\ \hline
    12 & 148 & 72 & 0.308 \\ \hline
		13 & 160 & 78 & 0.310 \\ \hline
    14 & 172 & 84 & 0.312 \\ \hline
    15 & 184 & 90 & 0.313 \\ \hline
		16 & 196 & 96 & 0.314 \\ \hline
    17 & 208 & 102 & 0.315 \\ \hline
    18 & 220 & 108 & 0.316 \\ \hline
		19 & 232 & 114 & 0.317 \\ \hline
    20 & 244 & 120 & 0.318 \\ \hline		
\end{tabular}
\end{table}

\begin{table}
\centering
\caption{Data for the calculation of CTR using new algorithm}
\label{tab:t2}
\begin{tabular}{|l|l|l|l|l|}
    \hline
    Time & Impressions & Clicks & Total Clicks & CTR\\ \hline
    1 & 16 & 2 & 22 & 0.090 \\ \hline
    2 & 28 & 6 & 50 & 0.12 \\ \hline
    3 & 40 & 12 & 84 & 0.142 \\ \hline
		4 & 52 & 18 & 124 & 0.145 \\ \hline
    5 & 64 & 24 & 170 & 0.141 \\ \hline
    6 & 76 & 30 & 222 & 0.135 \\ \hline
		7 & 88 & 42 & 280 & 0.128  \\ \hline
    8 & 100 & 48 & 344 & 0.122 \\ \hline
    9 & 112 & 54 & 414 & 0.115 \\ \hline
		10 & 124 & 60 & 490 & 0.110 \\ \hline
    11 & 136 & 66 & 572 & 0.104 \\ \hline
    12 & 148 & 72 & 660 & 0.100 \\ \hline
		13 & 160 & 78 & 754 &0.095 \\ \hline
    14 & 172 & 84 & 854 & 0.091\\ \hline
    15 & 184 & 90 & 960 & 0.087\\ \hline
		16 & 196 & 96 & 1072 & 0.083\\ \hline
    17 & 208 & 102 & 1190 & 0.080\\ \hline
    18 & 220 & 108 & 1314 & 0.077\\ \hline
		19 & 232 & 114 & 0.317 & 0.074\\ \hline
    20 & 244 & 120 & 1444 & 0.072\\ \hline		
\end{tabular}
\end{table}

\begin{figure}[bpht!]
\centering
\begin{subfigure}{.5\textwidth}
  \centering
  \includegraphics[width=\linewidth]{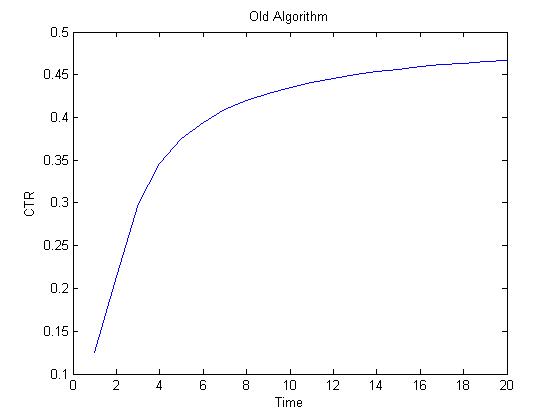}
  \caption{CTR in case of fraudulent clicks with old algorithm}
  \label{fig:sub1}
\end{subfigure}%
\begin{subfigure}{.5\textwidth}
  \centering
  \includegraphics[width=\linewidth]{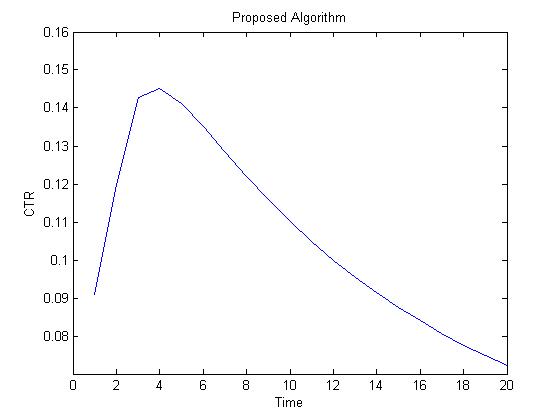}
  \caption{CTR in case of fraudulent clicks with proposed algorithm}
  \label{fig:sub2}
\end{subfigure}
\caption{CTR in case of fraudulent click with old and proposed algorithm}
\label{fig:test}
\end{figure}

Table \ref{tab:t1} depicts the CTR which is calculated by the current algorithm in the case of fraudulent clicks. Figure~\ref{fig:sub1} shows the exponential variation of CTR with time in case of fraudulent clicks using old algorithm. This problem is overcome by our proposed algorithm, which uses the same data for clicks but it raises up to a certain level till the normal clicks take place but eventually goes down when there are too many clicks within a very short duration of time. Thus the problem of rise in CTR due to fraudulent clicks is solved which is illustrated in Fig.~\ref{fig:sub2}.
\section{Conclusion} 
In this paper we have discussed the working of sponsored search and the business model of search engine based on sponsored search. Also we have discussed about the problem associated with the existing algorithm and shown the efficacy of the proposed algorithm over the existing one.

%
\bibliographystyle{IEEEtran}
\bibliography{myref}

\end{document}